\documentclass[9pt,twocolumn,abstract=true]{scrartcl}
\usepackage{authblk}
\usepackage[detect-weight=true, detect-family=true]{siunitx}
\usepackage{physics}
\usepackage{graphicx}
\usepackage{microtype}
\usepackage{hyperref}
\usepackage[capitalise]{cleveref}

\title{Direct characterisation of tuneable few-femtosecond dispersive-wave pulses in the deep UV}

\author[1,2,*]{\large{Christian Brahms}}
\author[1]{Dane R. Austin}
\author[3]{Francesco Tani}
\author[1,$\dagger$]{Allan S. Johnson}
\author[1]{Douglas Garratt}
\author[2,3]{John C. Travers}
\author[1]{John W.G. Tisch}
\author[3]{Philip St.J. Russell}
\author[1]{Jon P. Marangos}

\affil[1]{\small{Blackett Laboratory, Imperial College London, Prince Consort Road, London SW7 2AZ, UK}}
\affil[2]{School of Engineering and Physical Sciences, Heriot-Watt University, Edinburgh, EH14 4AS, UK}
\affil[3]{Max Planck Institute for the Science of Light, Staudtstr. 2, 91058 Erlangen, Germany}

\affil[$\dagger$]{Current address: ICFO - The Institute of Photonic Sciences, 08860 Castelldefels (Barcelona), Spain}
\affil[*]{Corresponding author: c.brahms@hw.ac.uk}

\begin{document}
\maketitle

\textbf{
Dispersive wave emission (DWE) in gas-filled hollow-core dielectric waveguides is a promising source of tuneable coherent and broadband radiation, but so far the generation of few-femtosecond pulses using this technique has not been demonstrated. Using in-vacuum frequency-resolved optical gating, we directly characterise tuneable \SI{3}{\fs} pulses in the deep ultraviolet generated via DWE. Through numerical simulations, we identify that the use of a pressure gradient in the waveguide is critical for the generation of short pulses.
}
	

Sources of very short laser pulses in the deep ultraviolet (DUV, \SIrange{200}{300}{\nm}) are a key enabling technology for many areas of physics, such as ultrafast spectroscopy \cite{stolow_femtosecond_2004, hockett_time-resolved_2011}. New approaches are made necessary by the limitations of sources based on harmonic generation, most importantly the lack of direct spectral tuneability, low conversion efficiency to the DUV, and the fact that only pulses of similar duration to the driving field can be generated \cite{reiter_generation_2010}. One promising avenue is the use of dispersive wave emission (DWE), a phenomenon observable during optical soliton \cite{wai_nonlinear_1986} and filamentation \cite{loures_superresonant_2017} dynamics among other effects. The extreme spectral broadening and nonlinear phase evolution experienced by a pulse undergoing soliton self-compression allows for phase-matching and coherent energy transfer to a secondary pulse at a different frequency, known as a dispersive wave. Crucially, in a gas-filled waveguide the wavelength at which phase-matching occurs is continuously tuneable by way of the gas pressure, enabling the generation of broadband pulses from the visible to the vacuum ultraviolet spectral range \cite{joly_bright_2011,mak_tunable_2013}.

Numerical studies suggest that the dispersive wave is generated as a near transform-limited pulse with a duration of only a few femtoseconds, significantly shorter than the driving pulse \cite{travers_ultrafast_2011}. Measurements of pulses which were strongly chirped by propagation through windows and ambient air before characterisation were consistent with few-fs pulse durations at generation, providing some support for this idea \cite{ermolov_characterization_2016,kottig_notitle_2015}. However, to date, no experiment has successfully measured a few-fs dispersive wave in the DUV or provided a route towards delivering such pulses to experiments without distortions.

Here we present the full characterisation of \SI{3}{\fs} dispersive-wave pulses with central wavelength tuneable from \SI{225}{\nm} to \SI{300}{\nm} using cross-correlation frequency-resolved optical gating (XFROG). Our measurements represent the first characterisation of a few-fs pulse generated via DWE and the shortest tuneable laser pulses in this spectral region to date, demonstrating the power of DWE-based sources for ultrafast science. The pulses are delivered to the characterisation set-up in vacuum; the measurements therefore faithfully reflect the pulses as generated in the waveguide. By numerically simulating the generation process, we further establish that the use of a pressure gradient along the waveguide is critical to the generation of few-fs pulses.

\begin{figure}[t]
	\centering
	\includegraphics[width=\linewidth]{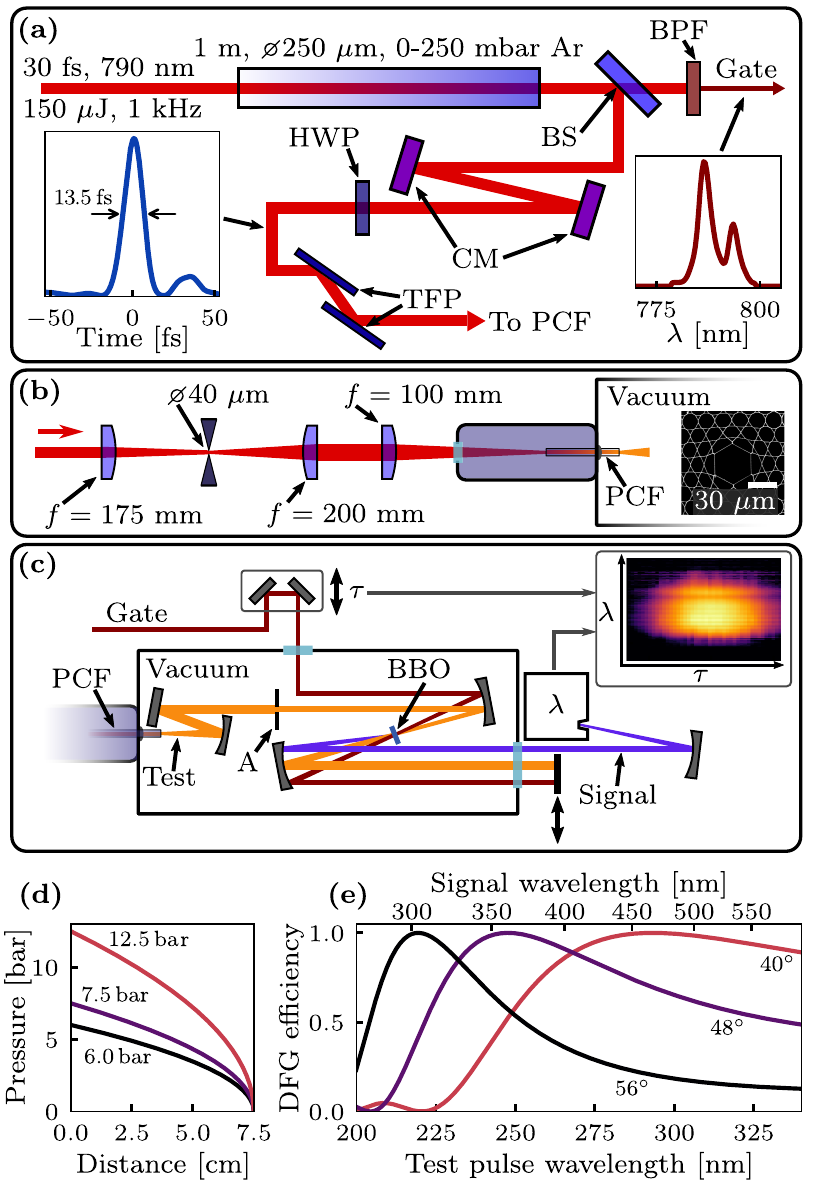}
	\caption{Layout of the experimental apparatus for (a) pre-compression, (b) DUV generation and (c) DUV pulse characterisation. BPF: bandpass filter; BS: beam-splitter; CM: chirped mirror; HWP: half-wave plate; TFP: thin-film polariser; $\tau$: delay stage, $\lambda$: spectrometer, A: aperture. (d) Pressure distribution along the \SI{7.5}{\cm} long AR-PCF for the input fill pressures used in the experiment. (e) Phase-matching efficiency for DFG between a DUV test pulse and a narrowband gate pulse centred at \SI{790}{\nm} for a \SI{5}{\micro\meter} BBO crystal for three different angles between the crystal axis and the beam propagation direction, taking into account the change in effective thickness as the crystal, cut for an angle of \ang{48}, is rotated away from normal incidence.}
	\label{fig:1}
\end{figure}

The experimental layout is shown in Fig.~\hyperref[fig:1]{1(a)-(c)} and consists of three parts: pre-compression, DUV generation, and pulse characterisation. A titanium-doped sapphire laser amplifier delivers pulses of \SI{30}{\fs} duration and \SI{150}{\micro\joule} energy at a repetition rate of \SI{1}{\kilo\hertz}. The pulses are spectrally broadened in a \SI{1}{\meter} long hollow glass capillary with a \SI{250}{\micro\meter} core diameter filled with a positive pressure gradient from vacuum to \SI{250}{\milli\bar} of argon and pass through a chirped-mirror compressor and an attenuator. To generate DUV pulses, the pre-compressed driving pulses are then coupled into a \SI{7.5}{\cm} long kagom\'{e}-type anti-resonant photonic crystal fibre (AR-PCF) with a core diameter of \SI{33}{\micro\meter} by focusing the beam to a \SI{22}{\micro\meter} diameter ($1/\mathrm{e}^2$) spot. A cross-section micrograph of the PCF is shown in the inset in Fig.~\hyperref[fig:1]{1(b)}. A spatial filter consisting of a \SI{40}{\micro\meter} diameter pinhole placed in the focal plane of a telescope removes distortions from the beam profile to optimise the coupling. The inset in Fig.~\hyperref[fig:1]{1(a)} shows the pulse profile at the entrance of the AR-PCF as measured using a second-harmonic generation (SHG) FROG device. Pre-compression of the driving pulse is not required but leads to higher-quality soliton self-compression and less structure in the DWE spectrum \cite{travers_ultrafast_2011}.

The AR-PCF is sealed into the connection between a gas cell and a vacuum chamber such that its entrance can be pressurised while keeping the exit under vacuum. This is critical to avoid stretching of the DUV pulse in an exit gas cell or a window. The shape of the resulting pressure gradient is given by $P(z) = P_0\sqrt{(L-z)/L}$, where $z$ is the distance along the waveguide, $L$ is the length of the waveguide and $P_0$ is the fill pressure at the entrance \cite{mak_tunable_2013}. Fig.~\hyperref[fig:1]{1(d)} shows $P(z)$ in the AR-PCF used in the experiment for three different fill pressures.

Due to the constraints of the vacuum system, DUV pulse energy measurement and pulse characterisation cannot be carried out simultaneously. An energy measurement with the gas cell at \SI{7.5}{\bar} pressure of argon and the vacuum chamber at atmospheric pressure showed a DUV pulse energy of \SI{30}{\nano\joule} with \SI{1}{\micro\joule} driving pulse energy, corresponding to a conversion efficiency of \SI{3}{\percent}. This is in line with previous results \cite{mak_tunable_2013} but can likely be optimised further. Higher energies can be achieved by moving to a larger core size \cite{kottig_generation_2017}.

The pulse characterisation is based on in-vacuum XFROG between a DUV pulse (the test pulse) and a narrowband infrared pulse (the gate) in a beta barium borate (BBO) crystal (see Fig.~\hyperref[fig:1]{1(c)}) \cite{linden_amplitude_1999}. In contrast to gas-based autocorrelation measurements \cite{reiter_generation_2010}, XFROG allows for the full characterisation of the temporal intensity and phase of the pulses. We use difference-frequency generation (DFG) as the nonlinear interaction since BBO is opaque below \SI{190}{\nm}, precluding the use of sum-frequency generation (SFG). Compared to third-order processes, DFG has the advantage of higher sensitivity as well as generating a signal at a different wavelength to the test pulse; this removes a potential source of background from the measurement \cite{linden_amplitude_1999}. As shown in Fig.~\hyperref[fig:1]{1(e)}, using a type I (o-o-e) scheme in a \SI{5}{\micro\meter} thick BBO crystal, broadband phase-matching for DFG across the DUV can be achieved by tuning the crystal angle.

A portion of the pulse is split off after the hollow capillary and bandpass-filtered to create the gate pulse; its spectrum is shown in the inset in Fig.~\hyperref[fig:1]{1(a)}. It enters the vacuum chamber through a separate window after traversing a motorised delay line. An aperture in the test pulse arm passes the DUV part of the spectrum undisturbed while removing over \SI{70}{\percent} of the driving pulse energy. This prevents nonlinear interactions between the driving pulse and the DUV pulse. The gate and test pulses are brought to a common focus in the BBO crystal, and the three beams exiting the crystal are collimated and steered out of the chamber for analysis by a spectrometer. The signal spot is imaged onto the spectrometer entrance slit with sufficient magnification to reduce the temporal blurring introduced by the non-collinear geometry to much less than the delay step in the XFROG trace. A moveable block allows for the measurement of the test pulse spectrum in the same position.

Pulse retrieval from XFROG measurements has traditionally relied upon the accurate characterisation of the gate pulse \cite{linden_amplitude_1999}. Recently, the use of ptychographic techniques has enabled the retrieval of pulses from XFROG traces without a known gate pulse \cite{witting_time-domain_2016}. Furthermore, it was demonstrated that ptychographic XFROG is capable of measuring pulses many times shorter than the gate \cite{spangenberg_all-optical_2016}. We apply this method to DFG XFROG for the first time. We use the regularised iterative ptychographic engine (rPIE) outlined in \cite{maiden_further_2017} with the addition of coefficient randomisation and soft thresholding described in \cite{sidorenko_ptychographic_2016} to aid convergence. To transfer the algorithm from SFG to DFG it has to be adapted, because the signal field in DFG XFROG is given by
\begin{equation}
	\psi(t) = E(t)G^*(t-\tau)\,,
	\label{eq:psi_DFG}
\end{equation}
where $E(t)$ is the field of the test pulse and $G(t- \tau)$ is the field of the gate pulse delayed by $\tau$, whereas in the case of SFG XFROG, the gate pulse appears without the complex conjugate. The adapted algorithm is thus derived by replacing the gate with its complex conjugate in the SFG algorithm. The rPIE update function to obtain a new guess for the test pulse, $E'(t)$, is given by
\begin{equation}
	E'(t) = E(t) + \frac{G(t-\tau_j)}{(1-\alpha)\abs*{G(t-\tau_j)}^2 + \alpha\abs*{G(t-\tau_j)}^2_{\mathrm{max}}}\Delta\psi(t)\,,
	\label{eq:test_update}
\end{equation}
where $\alpha$ is a randomly chosen coefficient of order unity, $_{\mathrm{max}}$ denotes the maximum value, and $\Delta\psi$ is given by $\Delta\psi = \psi'(t) - \psi(t)$; $\psi$ is the signal field as calculated with \cref{eq:psi_DFG} using the current guess for $E(t)$ and $G(t)$ and $\psi'$ is obtained by replacing the magnitude of $\psi$ with that of the measured XFROG trace:
\begin{equation}
	\psi'(t) = \mathcal{F}_{\omega \rightarrow t}^{-1}\left[
		\frac{
			\mathcal{F}_{t \rightarrow \omega}\left[\psi(t)\right]
		}{
			\abs{\mathcal{F}_{t \rightarrow \omega}\left[\psi(t)\right]}
		}\sqrt{S(\omega, \tau_j)}\right]
	\,.
\end{equation}
Here $\mathcal{F}_{t \rightarrow \omega}$ and $\mathcal{F}_{\omega \rightarrow t}^{-1}$ denote the forward and inverse Fourier transform over time, respectively, and $S(\omega, \tau_j)$ is the slice of the trace measured at delay $\tau_j$.
The gate pulse, on the other hand, is updated by
\begin{equation}
	G'(t-\tau_j) = G(t-\tau_j) + \frac{E(t)}{(1-\beta)\abs{E(t)}^2 + \beta\abs{E(t)}^2_{\mathrm{max}}}\Delta\psi^*(t)\,,
	\label{eq:gate_update}
\end{equation}
where $\beta$ is a second coefficient, and optionally projected onto the separately measured power spectrum of the gate pulse, $I_\textsc{g}(\omega)$:
\begin{equation}
	G''(t) = \mathcal{F}_{\omega \rightarrow t}^{-1} \left[
		\frac{
			\mathcal{F}_{t \rightarrow \omega}\left[G'(t)\right]}
		{
			\abs{\mathcal{F}_{t \rightarrow \omega}\left[G'(t)\right]}
		}
		\sqrt{I_\textsc{g}(\omega)}
		\right]\,.
\end{equation}

\begin{figure}[t]
	\centering
	\includegraphics[width=\linewidth]{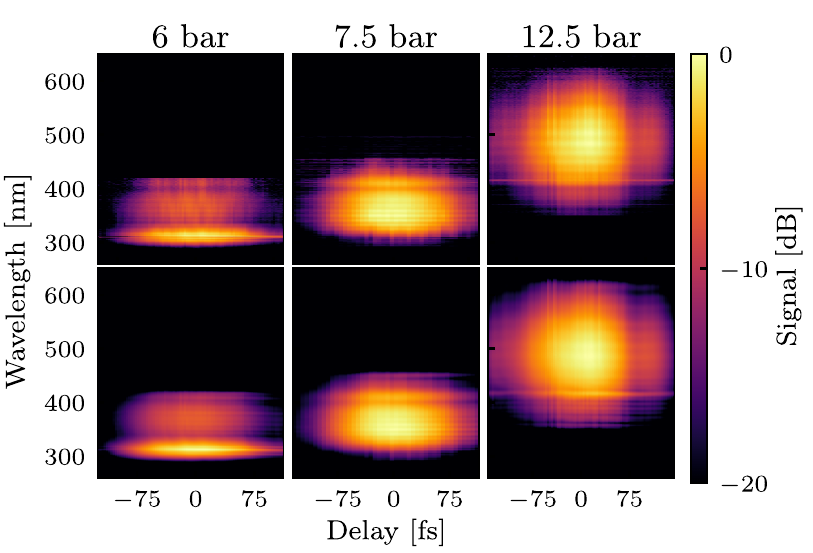}
	\caption{Measured (top row) and reconstructed (bottom row) DFG XFROG traces on a logarithmic colour scale for three different argon gas pressures at the PCF entrance. The measured traces have been corrected for the calculated phase-matching efficiency.}
	\label{fig:2}
\end{figure}

\begin{figure}[t]
	\centering
	\includegraphics[width=\linewidth]{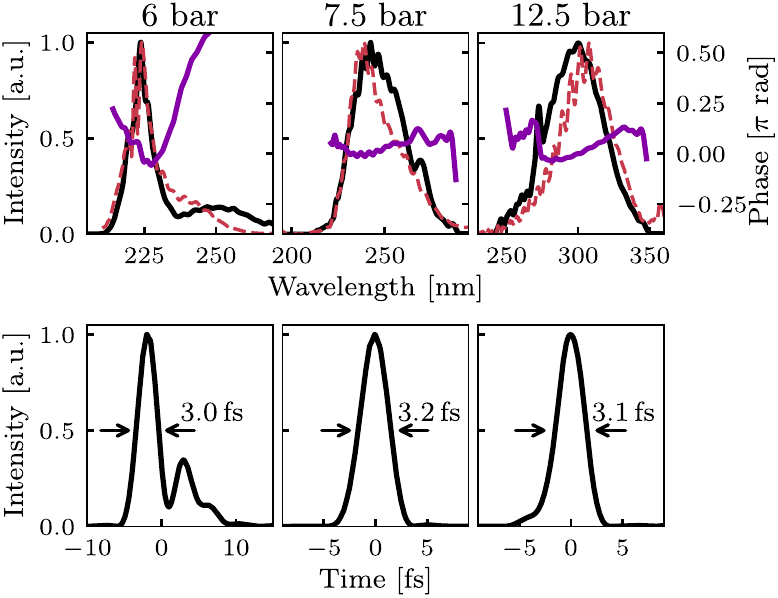}
	\caption{Spectral (top row) and time-domain (bottom row) representation of the DUV pulses as retrieved from the measurements shown in \cref{fig:2}. In the frequency domain plots, the black line shows the intensity, the purple line shows the spectral phase, and the red dashed line shows the separately measured power spectrum of the test pulse.}
	\label{fig:3}
\end{figure}

XFROG measurements of DUV pulses generated by coupling driving pulses with between \SI{0.65}{\micro\joule} and \SI{1.15}{\micro\joule} of energy into the AR-PCF filled with three different pressures of argon at the entrance are shown in \cref{fig:2}. The traces were corrected for the calculated phase-matching efficiency to account for the fact that, while broadband, it is not uniform across the bandwidth of the DUV pulses. Even before retrieval, it can be seen that the pulses are close to transform-limited by the absence of any tilt in the XFROG traces. The retrieval algorithm was run for 400 iterations with the spectral projection of the gate pulse applied for the first 50. The excellent agreement between the measured and retrieved traces shows that the retrieval was successful. The retrieved pulses are shown in \cref{fig:3} along with the spectrum of the pulses as measured by partially removing the block for the test and gate pulses such that the test pulse enters the spectrometer (see Fig.~\hyperref[fig:1]{1(c)}). The good agreement between the spectra confirms that the pulses were retrieved accurately.

\begin{figure}[t]
	\centering
	\includegraphics[width=\linewidth]{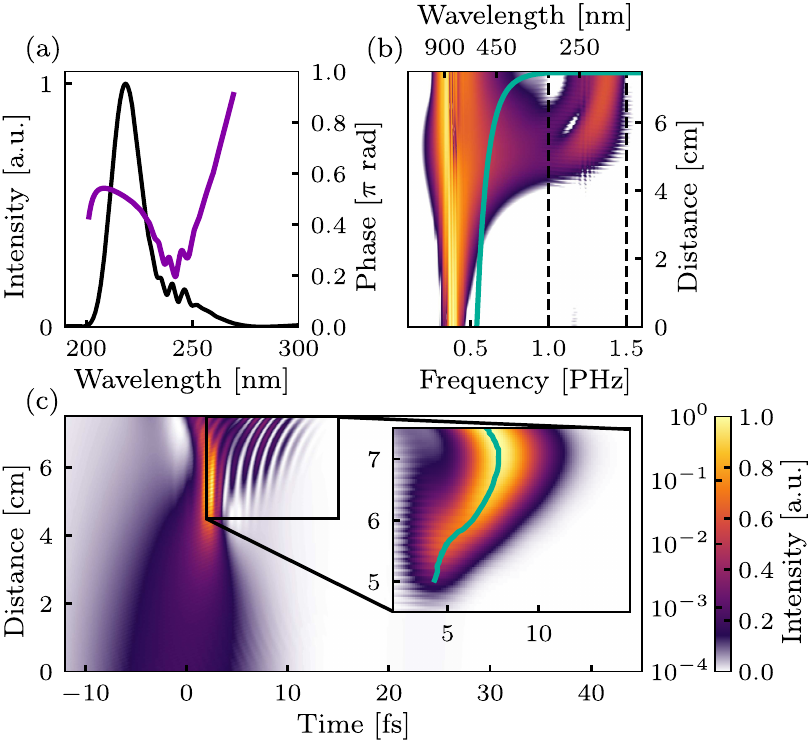}
	\caption{Simulated propagation of the measured input pulse with \SI{1.15}{\micro\joule} energy in a \SI{7.5}{\cm} length, \SI{33}{\micro\meter} core diameter AR-PCF filled with argon with a negative pressure gradient from \SI{6}{\bar} to vacuum (identical parameters to the experiment). (a) The output DUV spectral intensity (black) and phase (purple). (b) The spectral evolution of the pulse along the fibre on a logarithmic colour scale. The green line shows the zero-dispersion wavelength. (c) The temporal evolution of the field in a co-moving frame on a linear colour scale. The inset shows the evolution of the DUV part of the field as obtained by bandpass-filtering; the edges of the filter are shown as black dashed lines in (b). The green line follows the peak of the DUV pulse.}
	\label{fig:4}
\end{figure}

All of the measured pulses consist of a main peak with a duration of around \SI{3}{\fs}. To our knowledge, these are the shortest DUV pulses ever measured, with the exception of (non-tuneable and low-efficiency) third-harmonic generation using a few-cycle driving pulse \cite{reiter_generation_2010}. At \SI{250}{\nm} and \SI{300}{\nm} central wavelength, the pulse profile is exceptionally clean without post-pulses or significant pedestals. This is reflected in the flat spectral phase of these pulses, which confirms that the pulses are indeed very short at the point of generation and that our apparatus can deliver them to an experiment without significant dispersion. In contrast, the pulse at \SI{225}{\nm} exhibits a pedestal after the main peak. This is caused by the more complicated spectral intensity and phase structure, with a significant difference in arrival time between parts of the spectrum, as demonstrated by the strong phase gradient above \SI{235}{\nm}.

To better understand the conditions necessary for the generation of short DUV pulses, we have simulated the pulse propagation in the AR-PCF using the single-mode forward Maxwell equation \cite{couairon_practitioners_2011}. The model includes the waveguide \cite{marcatili_hollow_1964} and gas \cite{borzsonyi_dispersion_2008} dispersion, the Kerr nonlinearity \cite{shelton_measurements_1994}, and the effect of photoionisation \cite{geissler_light_1999} using the Perelomov-Popov-Terent'ev ionisation rate \cite{perelomov_ionization_1966}. The initial pulse energy was determined from the experimental energy and the transmission of the AR-PCF.

The DUV spectrum obtained by propagating the measured input pulse through a waveguide with the same parameters as used in the experiment at \SI{6}{\bar} argon pressure is shown in Fig.~\hyperref[fig:4]{4(a)}. The shape of the spectrum as well as the phase are in good agreement with the XFROG measurement, including a distinct difference in arrival time between the main peak and a long-wavelength pedestal. As shown in Fig.~\hyperref[fig:4]{4(b)}, the dispersive wave is emitted around \SI{5}{\cm} into the fibre. Due to the pressure gradient, the phase-matched wavelength changes over the course of the emission process, with lower pressures leading to phase-matching at shorter wavelengths \cite{mak_tunable_2013}. As a consequence the DUV pulse shifts and broadens spectrally as it is being generated, decreasing the transform-limited pulse duration of the dispersive wave. The lower-frequency pedestal appears further into the propagation and is displaced from the main pulse due to group-velocity mismatch. A result of complex generation dynamics, the pulse shape of the dispersive wave cannot be determined from its power spectrum alone. Accurate pulse characterisation is thus very important for the use of DWE as a practical source for ultrafast science.

In the time-domain representation, shown in Fig.~\hyperref[fig:4]{4(c)}, the self-compression of the driving pulse is clearly visible; DWE occurs at the point of maximum compression. The curved trajectory of the DUV pulse, revealed by bandpass filtering, shows that it begins to catch up to the driving pulse towards the end of the propagation after falling behind initially. This is due to the pressure gradient. As demonstrated by the shifting zero-dispersion wavelength, shown in Fig.~\hyperref[fig:4]{4(b)}, the dispersion landscape changes; the dispersive wave is accelerated relative to the driving pulse as a consequence. Importantly, despite several centimetres of propagation, the duration of the main DUV pulse remains well below \SI{4}{\femto\second}---the gas pressure is low and the DUV pulse experiences negligible dispersion. This is in contrast to the case of uniform gas pressure, where the dispersive wave can be stretched to several times its initial duration unless the point of maximum compression is very close to the end of the waveguide \cite{travers_ultrafast_2011}, making the generation process more sensitive to fluctuations of the input pulse energy. Since it also removes the need for an exit window as well as any propagation through a pressurised gas cell, the use of a pressure gradient is critical for the use of DWE as a source of few-fs pulses.

In summary, we have demonstrated the generation, delivery and characterisation of compressed, tuneable few-femtosecond pulses from DWE for the first time in what is also the first application of ptychographic retrieval to non-SFG XFROG. Our measurements as well as numerical simulations show that to obtain such short pulses, a pressure gradient in the waveguide has to be used. With vacuum propagation enabling the dispersion-free delivery of few-femtosecond DUV pulses, the apparatus is directly compatible with application in pump-probe schemes. Our work fully establishes DWE as a powerful light source for ultrafast measurements with unprecedented spectral coverage and temporal resolution.

\section*{Acknowledgements}
We acknowledge valuable assistance from T. Barnard, A. Ermolov, A. Gregory, S. Jarosch, E. Larsen, C. O'Donovan, S. Parker, C. Str\"{u}ber, and P. Ye, as well as funding from the following sources:
Engineering and Physical Sciences Research Council (EPSRC) (EP/I032517/1, EP/R019509/1);
European Research Council (ASTEX 290467, HISOL 679649);
EPSRC/Defence Science and Technology Laboratory Multidisciplinary University Research Initiative (EP/N018680/1);
EPSRC Doctoral Training Account

\bibliographystyle{unsrt}
\bibliography{references}

\end{document}